\documentclass[aps,prd,twocolumn,superscriptaddress,nofootinbib,floatfix]{revtex4-2}%
\usepackage[utf8]{inputenc}
\usepackage{amsmath,amssymb,bm}
\usepackage{graphicx}
\usepackage{booktabs}
\usepackage{xcolor}
\usepackage[colorlinks=true,linkcolor=blue!60!black,citecolor=blue!60!black,urlcolor=blue!60!black]{hyperref}

\newcommand{\dd}{{\rm d}}
\newcommand{\Veff}{V_{\rm eff}}
\newcommand{\VC}{V_{\rm C}}
\newcommand{\Vt}{\widetilde V}

\def\rf{\eqref}
\def\besub{\begin{subequations}}
\def\esub{\end{subequations}}

\def\db{\delta\beta}

\def\dpsi{\delta\psi}

\def\cL{{\cal L}}

\def\kappa{\varkappa}

\def\Veff{V_{\rm eff}}

\def\pb{perturbation}
\def\pbs{perturbations}
\def\sph{spherically symmetric}
\def\ssph{static, spherically symmetric}

\def\wh{wormhole}
\def\whs{wormholes}
\def\bh{black hole}
\def\bhs{black holes}

\def\Scw{Schwarz\-schild}
\def\Schr{Schr\"o\-din\-ger}

\usepackage{color}

\usepackage[normalem]{ulem}

\begin{document}
\tolerance 2000
%\usepackage{pyluatex}
%\begin{python}

%\if1\RevTeXmode
\title{Instabilities of Simpson--Visser black bounces sourced by a phantom scalar
and nonlinear electrodynamics}
\author{Kirill A. Bronnikov}
\email{kb20@yandex.ru}
\affiliation{Center of Gravitation and Fundamental Metrology, Rostest, Ozyornaya ul.\ 46, Moscow 119361, Russia}
\affiliation{Institute of Gravitation and Cosmology, RUDN University, ul.\ Miklukho-Maklaya 6, Moscow 117198, Russia}
\affiliation{National Research Nuclear University ``MEPhI'', Kashirskoe sh.\ 31, Moscow 115409, Russia}
\author{Rustam Ibadov}
\email{ibrustam@samdu.uz}
\affiliation{Department of Theoretical Physics and Quantum Electronics,
	 Institute of Engineering Physics, Samarkand State University, Samarkand 140104, Uzbekistan}
\author{Jutta Kunz}
\email{jutta.kunz@uni-oldenburg.de}
\affiliation{Institut f\"ur Physik, Universit\"at Oldenburg, Postfach 2503, D-26111 Oldenburg, Germany}
%\fi
%\if0\RevTeXmode
%\title{\bf Instabilities of Simpson--Visser black bounces sourced\\ by a phantom scalar and nonlinear electrodynamics}
%\author[1,2,3]{Kirill A. Bronnikov\thanks{kb20@yandex.ru}}
%\author[4]{Rustam Ibadov\thanks{ibrustam@samdu.uz}}
%\author[5]{Jutta Kunz\thanks{jutta.kunz@uni-oldenburg.de}}
%\affil[1]{Center of Gravitation and Fundamental Metrology, VNIIMS, Ozyornaya ul.\ 46, Moscow 119361, Russia}
%\affil[2]{Institute of Gravitation and Cosmology, RUDN University, ul.\ Miklukho-Maklaya 6, Moscow 117198, Russia}
%\affil[3]{National Research Nuclear University ``MEPhI'', Kashirskoe sh.\ 31, Moscow 115409, Russia}
%\affil[4]{Department of Theoretical Physics and Quantum Electronics, Institute of Engineering Physics, Samarkand State University, Samarkand 140104, Uzbekistan}
%\affil[5]{Institut f\"ur Physik, Universit\"at Oldenburg, Postfach 2503, D-26111 Oldenburg, Germany}
%\date{}
%\fi

\newcommand{\AbstractText}{%
    We study the linear stability of the Simpson--Visser (SV) family of ``black-bounce'' 
    space-times (regular \bhs\ and \whs) under spherically symmetric (monopole) perturbations,
    taking as a source a phantom scalar field $\psi$ with a self-interaction potential and a magnetic
    field within nonlinear electrodynamics (NED). The metric contains two parameters: the \Scw\ mass
    $M$ and the bounce parameter $a >0$, such that $a < 2M$ corresponds to the \bh\ branch, 
    and $a > 2M$ to \whs\ 
    (the extremal case $a=2M$ is not analyzed separately).
    Spherical perturbations possess a single dynamic degree of freedom, 
    and the stability problem reduces to a boundary-value problem for a Schr\"odinger-like 
    master equation with a certain effective potential $\Veff$. Our conclusion is that all space-times
    under study are unstable, and the corresponding exponential growth rates are calculated.   
    For the \bh\ branch the results are obtained without complications, while the \wh\ stability 
    problem faces difficulties that seem to be previously unnoticed. Specifically, the well-known 
    Darboux transformation, used to remove the wall-shaped $\Veff\approx 2/z^2$ in the effective   
    potentials at \wh\ throats ($z$ is the tortoise coordinate), creates nonphysical poles outside 
    the throat for the present \wh\ model if $a/(2M) \lesssim 4.98$. We circumvent this difficulty 
    by using a gauge-invariant approach in the spirit of Cremona, Pizzocchero and Sarbach, leading
    to a master equation with a simple and regular potential. Its further analysis reveals unstable 
    modes in all wormhole solutions considered, where the mode with a variable throat radius 
    is always present; for $a/(2M) \gtrsim 4.98$ there are two real growing throat modes, which
    merge into a complex-conjugate pair (an oscillatory instability) for smaller $a$.
    A more detailed spectral analysis reveals three exceptional points that separate 
    ranges of $a/(2M)$ with different perturbation regimes. Two of them are connected 
    by a common real eigenvalue branch. Furthermore, the leading unstable black-hole 
    and wormhole modes approach the same limiting growth rate 
    $M\kappa\simeq 0.10121$ as $a/(2M)$ approaches unity from either side.
    }

%\if1\RevTeXmode
\begin{abstract}
\AbstractText
\end{abstract}
\maketitle
%\fi
%\if0\RevTeXmode
%\twocolumn[{\maketitle\vspace{-1.2em}
%\begin{center}\begin{minipage}{0.86\textwidth}\small\noindent\textbf{Abstract.}
%\AbstractText\end{minipage}\end{center}\vspace{1.4em}}]
%\fi

%============================
\section{Introduction}\label{sec:intro}
%============================

  Black-bounce space-times \cite{SimpsonVisser2019}, the subject of the present paper, 
  unify in a single family regular \bhs\ and traversable Lorentzian wormholes. Both kinds of 
  objects have a long history: the former trace back to Bardeen's proposal in \cite{Bardeen1968}, 
  while the latter were obtained as exact solutions with a phantom scalar source in 
  \cite{Ellis1973} and \cite{Bronnikov1973}, and the whole concept was greatly popularized 
  by Morris and Thorne~\cite{MorrisThorne1988}. Some forms of nonlinear electrodynamics 
  (NED), other than its most well-known examples, the Born--Infeld and Heisenberg--Euler 
  theories \cite{BornInfeld1934, HeisenbergEuler}, have turned out to be a natural source of 
  regular \bhs\ \cite{AyonBeatoGarcia1998}, though solutions with a magnetic charge are 
  preferable compared to those with an electric charge \cite{kb-79, kb-01}.
  
  Simpson and Visser (SV) proposed a particularly economical geometric construction,
  the ``black-bounce'' metric~\cite{SimpsonVisser2019}, in which the \Scw\ metric is regularized
  by replacing the areal radius $r$ with $r(x)=\sqrt{x^2+a^2}$, introducing a single new length 
  scale $a$. Depending on the ratio $a/(2M)$, this space-time interpolates between the 
  Schwarzschild black hole ($a=0$), a regular black hole with a bounce hidden behind a horizon
  ($0<a<2M$), a regular extremal \bh\ ($a=2M$), and a traversable wormhole ($a>2M$).

  The SV proposal launched a large number of further studies. Its generalizations, their causal 
  structures and energy condition violations have been analyzed in \cite{LoboEtAl2021},
  and it is a standard item in surveys of regular black holes~\cite{LanEtAl2023}.
  Much recent work has focused on observational signatures at the test-field level:
  quasinormal ringing and echoes~\cite{YangEtAl2021}, and optical/shadow
  constraints~\cite{JafarzadeEtAl2022,AlencarEtAl2026Shadows}, the latter promoting
  massless wormhole seeds to massive black bounces and confronting their shadows and
  multiple light rings with the Event Horizon Telescope observations of Sgr~A*. The
  linear stability of more general black holes with NED sources is also actively   
  investigated~\cite{UktamovEtAl2026}. None of these studies, however, addresses the
  \emph{self-consistent} perturbation problem for the SV metric and its extensions. A reason is
  that the SV metric is originally purely geometric, while its \pb\ dynamics can only be analyzed 
  jointly with its possible matter sources. 

  In Ref.\ \cite{BronnikovWalia2022} it was noticed that any \ssph\ metric can be 
  obtained as a solution of general relativity with a combined source consisting of a scalar field 
  with a self-interaction potential and a magnetic field obeying NED, thus extending the 
  purely phantom constructions of \whs\ \cite{Ellis1973,Bronnikov1973} and regular black holes 
  \cite{BronnikovFabris2006}. In the general case, such a constituent scalar field could be not
  only purely canonical or phantom but also contain smooth transitions between such field 
  characteristics, forming the so-called ``trapped ghosts'' \cite{kb-sush-10}. For the case
  of SV black-bounce geometries (with or without a charge), such a field source is obtained in 
  \cite{BronnikovWalia2022} with a phantom field only. Examples of regularized metrics that 
  require partly phantom scalars (trapped ghosts) as sources were discussed in 
  \cite{kb-22-trap, kb-24-trap}.   
  
  All that concerned the possible existence of such objects themselves but not their stability.  
  Meanwhile, there have been extensive studies of static and rotating wormholes supported 
  by phantom fields, including their linear stability \cite{KleihausKunz2014, DzhunushalievEtAl2018, 
  ChewEtAl2019, GGSLinear2009, GGSNonlinear2009, GGSCharged2009,
  BronnikovFabrisZhidenko2011, BronnikovKonoplyaZhidenko2012}, 
  as well as \whs\ containing Yang--Mills fields~\cite{HauserEtAl2014} and
  those supported by higher-curvature terms instead of phantom matter
  \cite{IbadovEtAl2020EsGB, IbadovEtAl2022ArabJMath, IbadovEtAl2021Symmetry}.
  In particular, wormholes in dilatonic Einstein--Gauss--Bonnet gravity were first constructed
in Ref.~\cite{Kanti2011}, and studied in detail in Ref.~\cite{Kanti2012}. Subsequently they were
generalized to other scalar--Gauss--Bonnet coupling
functions~\cite{Antoniou2020}.
The radial stability of the dilatonic wormholes was
reexamined in Ref.~\cite{Cuyubamba2018}, where
unstable modes were found.
Further studies of Ellis--Bronnikov wormhole
perturbations have addressed polar modes and
isospectrality~\cite{Azad2023Polar}, as well as the
effect of slow rotation on the radial instability
\cite{Azad2024SlowRotation}.
  The stability of spherical systems involving scalar fields and NED was recently studied in 
  \cite{Ibadov2026, kb-NED-mirror, IbadovShukurullokhon2026, BronnikovEtAlBransDicke2025}.
  
  The present paper is devoted to analyzing the linear stability of the \Scw-SV construction 
  sourced by a phantom scalar and NED. As in a number of previous studies, while
  reducing the stability issue to spectral problems for a \Schr-like equation, we are here facing 
  a long-standing problem with the stability of throat-bearing configurations supported by 
  phantom scalars. The point is that the effective potential $\Veff$ for \sph\ \pbs\ of such 
  objects, obtained using the simplest \pb\ gauge $\delta r=0$ (or equivalently $\db =0$,
  see the metric \rf{eq:metric-gen}), as well as another gauge $\dpsi =0$,
  contains a wall shaped $\Veff \approx 2/z^2$ at the throat $z=0$ ($z$ being the tortoise
  coordinate), and this wall makes impossible considering the most important \pb\ mode 
  related to an evolving throat radius.
    
  This problem seemed to be solved by applying the S-deformation (Darboux) technique
  as suggested by Gonz\'alez, Guzm\'an and Sarbach in \cite{GGSLinear2009} who proved 
  the instability of massless phantom-scalar (Ellis-type) wormholes as well as their 
  electrically charged counterparts \cite{GGSCharged2009}. This approach was
  extended in \cite{BronnikovFabrisZhidenko2011} and applied in   
  \cite{BronnikovKonoplyaZhidenko2012}, where it was shown that almost all wormholes and
  regular black holes supported by a phantom scalar are unstable against spherical
  perturbations due to the restored throat-radius mode; magnetic wormholes and black 
  universes with phantom scalars were studied in~\cite{BronnikovKorolyov2015}, see also a 
  review in \cite{Bronnikov2018Particles}. 
  
  However, as we further discuss in the present paper, this approach relies on an essential 
  condition: isospectrality of the Darboux transformation for wave or \Schr-like
  equations holds only if the transformed potential is regular, i.e., if a nodeless seed
  function exists. It turns out that, 
  being applied to  $\Veff$ with a pole like $\approx 2/z^2$, the transformation removes this
  pole, but the new potential may contain nonphysical walls emerging outside the throat, 
  thus completely distorting the spectrum, and this really happens to the system presently 
  under study for $a/(2M)\lesssim 4.98$.
  
  We see a way out in the gauge-invariant formalism developed by Cremona, Pizzocchero and 
  Sarbach (CPS) in \cite{CPS2020} that produces a throat-regular master equation
  for wormholes with a self-interacting phantom field without any S-deformation.  Its more 
  general version involving NED is developed and applied here, see also \cite{Ibadov2026}.
  An alternative approach is to solve a coupled system of perturbation equations directly, without
reducing it to a Schr\"odinger-type equation with a singular potential at the throat.
This strategy has been employed for static Ellis--Bronnikov wormholes~\cite{BlazquezChewKunz2018} and, with a throat-regular system, for their charged
generalizations~\cite{BlazquezCharged2026}.

  Exceptional points (EPs), characterized by the coalescence of eigenvalues and eigenfunctions,
have recently been studied in black-hole quasinormal-mode spectra
\cite{CavalcanteEP2024,OshitaEP2025,YangEP2025}.
Their implications for resonant mode excitation and black-hole ringdown have also been
 investigated \cite{MotohashiEP2025,MacedoEP2026}.
More recently, an EP involving two dynamically unstable radial modes was identified for rotating
five-dimensional Ellis--Bronnikov wormholes \cite{EB5DEP2026}. Related mergers of unstable
modes had already been observed for charged Ellis--Bronnikov wormholes
\cite{BlazquezCharged2026}.

  In what follows we carry out a complete linear radial stability analysis of electrically neutral
  SV-\Scw\ configurations with a scalar-NED source \cite{BronnikovWalia2022}. 
   Our main points are:
\begin{itemize}
\item 
	In the gauges $\delta\beta=0$ and $\delta\psi=0$, the effective potential $\Veff$ is found 
	in a single closed form; it contains a universal wall with $\approx 2/z^2$ at the throat. 
	However, the physical conditions applied to admissible solutions are gauge-dependent, 
	therefore, the sets of solutions obtained in the two gauges are different.
\item 
	The Darboux transformation uses an auxiliary ``seed'' function $Y_s$, being a solution to the 
	master equation (in particular, one describing static \pbs). 
	A regular potential in the whole range of the radial
	coordinate is only obtained if such a seed is nodeless, $Y_s\ne 0$.
	We show that for SV wormholes with $M>0$ nodeless seeds are only obtained 
	if $a/(2M)\gtrsim 4.98$.
\item 
	A CPS-type gauge-invariant equation, with its simple regular potential
      $\VC=-6MAx^2/r^5$, solves the problem 
      for all $a \ne 2M$ and leads to the conclusion that 
      all wormholes and all black holes of this family are unstable. 
      The results are confirmed by time-domain evolution. 
\item
	A more detailed analysis of the unstable spectrum
reveals three exceptional points, two of which are connected by a common real eigenvalue branch.
Moreover, the leading unstable modes of the black hole
and wormhole branches approach the same growth rate
$M\kappa\simeq 0.10121$ as $a/(2M)$ approaches unity from either side.
\end{itemize}

  The paper is organized as follows. Section~\ref{sec:basic} gives the action, the field
   equations and the background. Section~\ref{sec:pert} discusses the perturbation equations
   in the gauges $\delta\beta=0$ and $\delta\psi=0$ and the black hole branch.
   Section\,\ref{sec:darboux} is devoted to Darboux transformations and isospectrality, 
   and Sec.\,\ref{sec:cps} to the CPS method, the exceptional point structure, 
   and a comparison of all results.
   Section\,\ref{sec:disc} contains a discussion.

%======================================================================
\section{Basic equations}  \label{sec:basic}
%======================================================================
\subsection{Action, metric and notations}
% ---------------------------------------------

We consider general relativity with a minimally coupled scalar field $\psi$ and NED,
\begin{align}
  S=\frac{1}{16\pi}\int\dd^4x\sqrt{-g}\,\Big[&R+2h(\psi)\,g^{\mu\nu}\psi_{,\mu}\psi_{,\nu}
  \nonumber\\ &-2V(\psi)-\mathcal{L}(\mathcal{F})\Big],
  \label{eq:action}
\end{align}
where $\mathcal F=F_{\mu\nu}F^{\mu\nu}$, $V(\psi)$ is a self-interaction potential and
$h(\psi)$ a kinetic coupling; $h=+1$ for a canonical and $h=-1$ for a phantom field.
We use the units $c=G=1$, the signature $(+,-,-,-)$ and the sign conventions
of~\cite{Bronnikov2018Particles}, for which the field equations read
\begin{equation}
  G^\nu_\mu\equiv R^\nu_\mu-\tfrac12\delta^\nu_\mu R=-T^\nu_\mu, \qquad
  T^\nu_\mu=T^\nu_\mu[\psi]+T^\nu_\mu[F],
  \label{eq:EE}
\end{equation}
with
\begin{align}
  T^\nu_\mu[\psi]&=2h\psi_{,\mu}\psi^{,\nu}-\delta^\nu_\mu\big(h\psi^{,\alpha}\psi_{,\alpha}-V\big),
  \label{eq:Tpsi}\\
  T^\nu_\mu[F]&=-2\mathcal L_{\mathcal F}F_{\mu\alpha}F^{\nu\alpha}+\tfrac12\delta^\nu_\mu\mathcal L,
  \qquad \mathcal L_{\mathcal F}\equiv \dd\mathcal L/\dd\mathcal F .
  \label{eq:TF}
\end{align}
The scalar field obeys
\begin{equation}
  2h\,\Box\psi+h_\psi\,\psi^{,\alpha}\psi_{,\alpha}+V_\psi=0,\qquad V_\psi\equiv \dd V/\dd\psi ,
  \label{eq:KG}
\end{equation}
and the NED equations are $\nabla_\mu(\mathcal L_{\mathcal F}F^{\mu\nu})=0$.

For time-dependent spherically symmetric configurations we write
\begin{equation}
  \dd s^2=e^{2\gamma}\dd t^2-e^{2\alpha}\dd u^2-e^{2\beta}\dd\Omega^2,
  \label{eq:metric-gen}
\end{equation}
with $\gamma,\alpha,\beta,\psi$ functions of $t$ and $u$, and $r\equiv e^\beta$ the
areal radius; $\dd\Omega^2=\dd\theta^2+\sin^2\theta\,\dd\varphi^2$. Dots and primes denote $\partial_t$ and $\partial_u$. The nonzero
components of the Einstein tensor, in the conventions of~\eqref{eq:EE}, are
\begin{align}
G^0_0&=-e^{-2\beta}+e^{-2\alpha}\big(2\beta''+3\beta'^2-2\alpha'\beta'\big)\nonumber\\
     &\quad -e^{-2\gamma}\big(\dot\beta^2+2\dot\alpha\dot\beta\big),\nonumber\\
G^1_1&=-e^{-2\beta}+e^{-2\alpha}\big(\beta'^2+2\beta'\gamma'\big)\nonumber\\
     &\quad -e^{-2\gamma}\big(2\ddot\beta+3\dot\beta^2-2\dot\gamma\dot\beta\big),\nonumber\\
G^2_2&=e^{-2\alpha}\big(\gamma''+\gamma'^2-\alpha'\gamma'+\beta''+\beta'^2-\alpha'\beta'
      +\beta'\gamma'\big)\nonumber\\
     &\quad -e^{-2\gamma}\big(\ddot\alpha+\dot\alpha^2-\dot\alpha\dot\gamma+\ddot\beta
      +\dot\beta^2-\dot\beta\dot\gamma+\dot\alpha\dot\beta\big)\nonumber\\
     &=G^3_3,\nonumber\\
G^1_0&=-2e^{-2\alpha}\big(\dot\beta'+\dot\beta\beta'-\dot\beta\gamma'-\dot\alpha\beta'\big).
\label{eq:Gmunu}
\end{align}

  For a radial magnetic field, $F_{23}=q\sin\theta$, the NED equations hold identically
  for any metric of the form~\eqref{eq:metric-gen}, be it static or time-dependent, so that
  $\mathcal F=2q^2e^{-4\beta}=2q^2/r^4$. Thus within any spherically symmetric configuration, 
  $\cL$ is a known function of $r$ alone, $\mathcal L(\mathcal F(r))\equiv L(r)$, and
\begin{align}
  &T^0_0[F]=T^1_1[F]=\tfrac12 L,\nonumber\\
  &T^2_2[F]=T^3_3[F]=\tfrac12 L-\mathcal F\mathcal L_{\mathcal F}=\tfrac12L+\tfrac14 rL_r,
  \label{eq:TFr}
\end{align}
  with $L_r=\dd L/\dd r$; the last equality follows from $\dd\mathcal F/\dd r=-4\mathcal F/r$.
  The charge $q$ enters only through the normalization of $\mathcal L(\mathcal F)$, and,
   once $L(r)$ is fixed by the background, it drops out from the whole problem.

% -----------------------------------
\subsection{Static equations}
% -----------------------------------

  For static configurations we use the quasiglobal gauge $\alpha+\gamma=0$ and denote
  $u\equiv x$, $A(x)\equiv e^{2\gamma}=e^{-2\alpha}$:
\begin{equation}
  \dd s^2=A(x)\dd t^2-\frac{\dd x^2}{A(x)}-r^2(x)\dd\Omega^2.
  \label{eq:metric-static}
\end{equation}
The independent field equations can be written as
\begin{align}
 & \frac{r''}{r}=-h\psi'^2, \label{eq:st1}\\
 & A(r^2)''-r^2A''=2-4q^2\mathcal L_{\mathcal F}/r^2, \label{eq:st2}\\
 & \frac{1}{r^2}\big(-1+A'rr'+Ar'^2\big)=hA\psi'^2-V-\tfrac12L, \label{eq:st3}\\
 & 2h\,(Ar^2\psi')'=r^2V_\psi \qquad (h={\rm const}), \label{eq:st4}
\end{align}
and Eq.~\eqref{eq:st4} follows from the others by the Bianchi identities.
Equation~\eqref{eq:st1} shows that a minimum of $r(x)$ (a throat or a bounce) requires
$h=-1$.

% -------------------------------------------------------
\subsection{Simpson--Visser metric and its source}
% -------------------------------------------------------

  The SV metric~\cite{SimpsonVisser2019} corresponds to
\begin{equation}
  r(x)=\sqrt{x^2+a^2},\quad\ A(x)=1-\frac{2M}{r(x)},\quad\ x\in\mathbb R,        \label{eq:SV}
\end{equation}
   with $M>0$ and $a>0$. For $a>2M$, $A>0$ everywhere, and the metric describes a symmetric
   traversable wormhole with the throat $r_{\rm th}=a$ at $x=0$; at $a<2M$ there are
   horizons at $x=\pm x_{\rm hor} =\sqrt{4M^2-a^2}$, and the bounce $x=0$ is hidden inside a
   T-region; at $a=2M$ we obtain a regular extremal \bh\ with a horizon at minimum radius, 
   $x_{\rm hor}=0$, $r_{\rm hor} =a$.

  Following \cite{BronnikovWalia2022}, Eqs.\,\eqref{eq:st1}--\eqref{eq:st4} with $h=-1$ and
  zero electric charge are solved by
\begin{align}
 &\psi(x)=\arctan\frac{x}{a},\quad V(\psi)=\frac{4M}{5a^3}\cos^5\psi
  =\frac{4Ma^2}{5r^5}, \label{eq:BWpsi}
\\[6pt]
 & L(r)=\frac{12Ma^2}{5r^5}\ \Longleftrightarrow\
  \mathcal L(\mathcal F)=\frac{12Ma^2}{5\,(2q^2)^{5/4}}\,\mathcal F^{5/4},   
  \label{eq:BWL}
\end{align}
  whence $\mathcal L_{\mathcal F}=3Ma^2/(2q^2r)$. One also has
  $\psi'=a/r^2$, $V_\psi=-4aMx/r^5$, and $V_{\psi\psi}=-4M(r^2-5x^2)/r^5$. Note that the 
  scalar field is monotonic, $\psi'>0$, on the whole real line, while
  $\beta'=r'/r=x/r^2$ vanishes at $x=0$; this difference underlies all that follows.
  The limit $M\to 0$ leads to the massless Ellis--Bronnikov wormhole, with $V=L=0$.

%======================================================================
\section{Perturbation equations}  \label{sec:pert}
%======================================================================

\subsection{Linearized equations and gauge freedom}

We consider small time-dependent spherically symmetric perturbations of a static
background~\eqref{eq:metric-static},
\begin{align}
  \gamma&=\gamma(x)+\delta\gamma(t,x),&\alpha&=\alpha(x)+\delta\alpha(t,x),\nonumber\\
  \beta&=\beta(x)+\delta\beta(t,x),&\psi&=\psi(x)+\delta\psi(t,x),
  \label{eq:pert}
\end{align}
with $e^{2\gamma(x)}=e^{-2\alpha(x)}=A(x)$ and $e^{\beta(x)}=r(x)$. The magnetic field
needs no perturbation: $F_{23}=q\sin\theta$ is exact, and its only effect is through
$L(r)$ in~\eqref{eq:TFr}, so that $\delta T^0_0[F]=\delta T^1_1[F]=\frac12L_\beta\,\delta\beta$
with $L_\beta\equiv \dd L/\dd\beta=rL_r$. In what follows $h=-1$.

Linearizing Eqs.~\eqref{eq:EE} and~\eqref{eq:KG} with~\eqref{eq:Gmunu}, \eqref{eq:Tpsi},
\eqref{eq:TFr}, we obtain the $\binom{1}{0}$ (momentum) constraint, integrated in $t$,
\begin{equation}
  \delta\beta'+(\beta'-\gamma')\delta\beta-\beta'\delta\alpha=\psi'\delta\psi,
  \label{eq:C01}
\end{equation}
the $\binom{1}{1}$ and $\binom{0}{0}$ components,
\begin{align}
 &\frac{2\delta\beta}{r^2}+2A\Big[(\beta'+\gamma')\delta\beta'+\beta'\delta\gamma'
  -\beta'(\beta'+2\gamma')\delta\alpha\Big]-\frac{2}{A}\delta\ddot\beta \nonumber\\
 &\ \qquad =-2A\big(\psi'\delta\psi'-\psi'^2\delta\alpha\big)%\nonumber\\
         -V_\psi\delta\psi-\tfrac12L_\beta\delta\beta,
  \label{eq:C11}\\
 &\frac{2\delta\beta}{r^2}+2A\Big[\delta\beta''+(3\beta'+\gamma')\delta\beta'-\beta'\delta\alpha'
  \nonumber\\
 &\qquad\qquad\qquad -(2\beta''+3\beta'^2+2\beta'\gamma')\delta\alpha\Big] \nonumber\\
 &\ \qquad =2A\big(\psi'\delta\psi'-\psi'^2\delta\alpha\big)%\nonumber\\
         -V_\psi\delta\psi-\tfrac12L_\beta\delta\beta,
  \label{eq:C00}
\end{align}
and the scalar field equation
\begin{align}
  &\frac{\delta\ddot\psi}{A}-\frac{1}{r^2}\big(Ar^2\delta\psi'\big)'
  -A\psi'\big(\delta\gamma'-\delta\alpha'+2\delta\beta'\big)\nonumber\\
  &\qquad -V_\psi\delta\alpha -\tfrac12V_{\psi\psi}\delta\psi=0.
  \label{eq:dKG}
\end{align}
The $\binom22$ component follows from these by the Bianchi identities.

The perturbations are defined up to infinitesimal coordinate transformations
$x\to x+\xi(t,x)$, $t\to t+\eta(t,x)$, under which
\begin{align}
 &\delta\beta\to\delta\beta+\beta'\xi,\qquad \delta\psi\to\delta\psi+\psi'\xi,\nonumber\\
 &\delta\alpha\to\delta\alpha+\alpha'\xi+\xi',\qquad \delta\gamma\to\delta\gamma+\gamma'\xi+\dot\eta.
 \label{eq:gauge}
\end{align}
Accordingly, there are two natural gauge-invariant combinations of $\delta\beta$ and
$\delta\psi$,
\begin{equation}
  \Phi\equiv\delta\psi-\frac{\psi'}{\beta'}\,\delta\beta,\quad
  \mathcal B\equiv\delta\beta-\frac{\beta'}{\psi'}\,\delta\psi,\quad
  \Phi=-\frac{\psi'}{\beta'}\,\mathcal B .
  \label{eq:GI}
\end{equation}
$\Phi$ coincides with $\delta\psi$ in the gauge $\delta\beta=0$, and $\mathcal B$ with
$\delta\beta$ in the gauge $\delta\psi=0$. For the SV background $\psi'/\beta'=a/x$,
so $\Phi$ is singular at the throat $x=0$, whereas $\mathcal B$ is regular everywhere,
because $\psi'=a/r^2$ never vanishes. The gauge $\delta\beta=0$ is attainable only for
perturbations with $\delta\beta=0$ at the throat, whereas the gauge $\delta\psi=0$ is
always attainable. This is the source of all subtleties discussed below.

% --------------------------------------------------------------------------------------
\subsection{Gauge $\delta\beta=0$: Master equation and a wall at the throat}
% --------------------------------------------------------------------------------------

\begin{figure*}[t]
\centering
\includegraphics[width=0.96\textwidth]{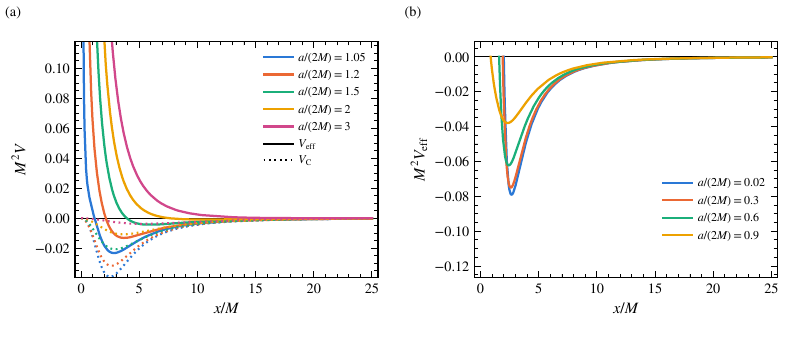}
\caption{(a) Wormhole branch: the master potential $\Veff$ of Eq.~\eqref{eq:Veff-SV}
(solid) with the wall $2/z^2$ at the throat, and the CPS potential
$\VC=-6MAx^2/r^5$ of Eq.~\eqref{eq:VC} (dotted, same colors), which is regular. Only
$x>0$ is shown; both are even in $x$. (b) Black-hole branch: $\Veff$ in the exterior
region $x>x_h$. Units $M=1$.}
\label{fig:pot}
\end{figure*}

In the gauge $\delta\beta=0$, Eq.~\eqref{eq:C01} gives
$\delta\alpha=-(\psi'/\beta')\delta\psi$, Eq.~\eqref{eq:C11} gives $\delta\gamma'$, and
Eq.~\eqref{eq:dKG} then becomes a closed wave equation. The NED field does not
contribute at all, since $\delta L=L_\beta\delta\beta=0$. The terms with
$\delta\psi'$ that arise from $\delta\gamma'$ and $\delta\alpha'$ cancel, and with
\begin{equation}
  \delta\psi=\frac{Y(x)}{r(x)}\,e^{i\Omega t},\qquad \dd z=\frac{\dd x}{A(x)},
  \label{eq:Ydef}
\end{equation}
we obtain the Schr\"odinger-type master equation
\begin{equation}
  \frac{\dd^2Y}{\dd z^2}+\big[\Omega^2-\Veff(z)\big]Y=0,
  \label{eq:master}
\end{equation}
with the effective potential (valid for any $V(\psi)$ and any magnetic NED with $h=-1$)
\begin{align}
\Veff={}&\frac{A(Ar')'}{r}+A^2\psi'^2\Big[1+\frac{rA'}{Ar'}+\frac{r^2\psi'^2}{r'^2}\Big]
  \nonumber\\ &-A^2\psi'\Big(\frac{r\psi'}{r'}\Big)'+\frac32A\frac{r\psi'}{r'}\,V_\psi-\frac12AV_{\psi\psi}.
\label{eq:Veff-gen}
\end{align}
For the SV background~\eqref{eq:SV}--\eqref{eq:BWL} this gives the closed form
\begin{align}
  \Veff(x)&=\frac{A(x)\,N(x)}{x^2r^5},\nonumber\\
  N(x)&=a^2r\,(2r^2+x^2)-2M\big(2a^4+4a^2x^2+3x^4\big),
  \label{eq:Veff-SV}
\end{align}
which depends on $M$ and $a$ but not on $q$. We have verified Eq.~\eqref{eq:Veff-SV}
against a numerical evaluation of~\eqref{eq:Veff-gen} and, independently, against the
equations in the gauge $\delta\psi=0$ (see below), including all NED terms. At large
$|x|$, $\Veff\simeq-6M/|x|^3$; for $M=0$ it reduces to the Ellis potential
$\Veff=2a^2/(x^2r^2)+a^2/r^4$. The potential is shown in Fig.~\ref{fig:pot}.
% ------------------------------------------------- fig 1
% -------------------------------------------------------------------

\emph{The wall.} Near the throat of a wormhole ($a>2M$), $A(0)=1-2M/a>0$,
$z\simeq x/A(0)$, and
\begin{equation}
  \Veff=\frac{2}{z^2}+w_0+w_2z^2+O(z^4),
  \label{eq:wall}
\end{equation}
independently of $M$ and $a$, in agreement with the general result for generic
throats~\cite{Bronnikov2018Particles}. The indicial equation $p(p-1)=2$ has the roots
$p=2$ and $p=-1$, which differ by an integer. Hence the solutions behave as
\begin{align}
  Y_{\rm reg}&=z^2\big(1+O(z^2)\big),\nonumber\\
  Y_{\rm sing}&=\frac1z\Big(1+\frac{\Omega^2-w_0}{2}\,z^2+O(z^4)\Big)+c_2\,Y_{\rm reg},
  \label{eq:frob}
\end{align}
and the coefficient $c_2$ is free (no logarithm is forced); $c_2=0$ selects the
solution that is odd in $z$. In this gauge
$\delta\alpha=-(\psi'/\beta')\delta\psi\approx-aY/(rx)$, so $Y_{\rm sing}$ leads to
$\delta\alpha\sim x^{-2}$, which seems unacceptable. Imposing $Y(0)=0$ on each side of
the throat then turns the problem into two decoupled half-line problems with a Dirichlet
wall; we call this the \emph{naive} boundary condition. As we shall now see, it discards
physically regular perturbations.

% --------------------------------------------------
\subsection{Gauge $\delta\psi=0$: the same potential, different admissible solutions}
% --------------------------------------------------

In the gauge $\delta\psi=0$, the perturbation $\mathcal B=\delta\beta$ is the basic
unknown. Equation~\eqref{eq:C01} gives
\begin{equation}
  \delta\alpha=\frac{\delta\beta'+(\beta'-\gamma')\delta\beta}{\beta'},
  \label{eq:dalpha-psi}
\end{equation}
the scalar equation~\eqref{eq:dKG} gives
\[
	\delta\gamma'=\delta\alpha'-2\delta\beta'-V_\psi\delta\alpha/(A\psi'), 
\]	
  and then Eq.~\eqref{eq:C11} yields a wave equation for $\delta\beta$ alone, while
Eq.~\eqref{eq:C00} is satisfied identically. Now NED contributes explicitly through the
term $\frac12L_\beta\delta\beta$, and $V_\psi$ enters in a different way; using the
background equations, all these terms combine into a single equation, which in
terms of
\begin{equation}
	 Y=-\frac{r\psi'}{\beta'}\,\mathcal B=-\frac{ar}{x}\,\delta\beta \qquad(=r\Phi)
 \label{eq:Y-B}
\end{equation}
takes \emph{exactly} the form~\eqref{eq:master} with the same $\Veff$. We have checked
this numerically: solutions of~\eqref{eq:master}, converted to $\delta\beta$
by~\eqref{eq:Y-B}, satisfy the $\delta\psi=0$ equations to the accuracy of the
finite-difference test ($\sim10^{-8}$), and Eq.~\eqref{eq:C00} to machine precision.
This is no accident: since $Y=r\Phi$ is gauge-invariant, and the Schr\"odinger normal
form of a second-order equation in a given variable $z$ is unique, \emph{the effective
potential is the same in both gauges, wall included}.

What changes is the class of admissible solutions. Near $x=0$ the equation for
$\mathcal B$ has the leading part $\mathcal B''-(2/x)\mathcal B'$, with indicial exponents
$0$ and $3$; by Eq.~\eqref{eq:Y-B}, they correspond to $Y_{\rm sing}$ and $Y_{\rm reg}$,
respectively:
\begin{align}
  \mathcal B_{\rm even}&=1+\frac{\Omega^2}{2A(0)^2}x^2+O(x^4)\ \leftrightarrow\ Y_{\rm sing},
  \nonumber\\
  \mathcal B_{\rm odd}&=x^3+O(x^5)\ \leftrightarrow\ Y_{\rm reg}.
  \label{eq:B-loc}
\end{align}
Both are smooth, so $x=0$ is an \emph{apparent} singularity of the $\mathcal B$
equation. Moreover, by~\eqref{eq:dalpha-psi} and the expression for $\delta\gamma'$,
$\delta\alpha$ and $\delta\gamma$ are regular for both of them. Thus, in the gauge
$\delta\psi=0$, \emph{every} local solution is a regular perturbation, and the singular
branch $Y\sim 1/z$ of the $\delta\beta=0$ gauge is nothing else than a perturbation with
$\delta r(t,0)=a\,\delta\beta(t,0)\neq0$, a change of the throat radius. The resonance
constant $c_2$ in~\eqref{eq:frob} is the freedom to add the odd solution
$\mathcal B_{\rm odd}$. The singularity of $\delta\alpha$ in the gauge $\delta\beta=0$ is
a pure gauge effect.

Since the background is symmetric under $x\to-x$, the perturbations split into two
sectors:
\begin{itemize}
\item the \emph{odd} sector ($\mathcal B$ odd, $Y$ even, $Y\sim z^2$): the throat radius is
unperturbed; this is exactly the naive problem with $Y(0)=0$;
\item the \emph{even} sector ($\mathcal B$ even, $\mathcal B(0)\ne0$, $Y\sim1/z$): the
throat-radius (or ``breathing'') modes, which are completely lost under the naive
boundary condition.
\end{itemize}
The even sector cannot be treated directly by Eq.~\eqref{eq:master}, since
$\int Y^2\dd z$ diverges at the throat: the problem in the variable $Y$ is not a standard
self-adjoint one. Sections~\ref{sec:darboux} and~\ref{sec:cps} describe two ways around
this obstacle.

\emph{Naive (odd-sector) results.} For completeness, the odd sector is a regular
Sturm--Liouville problem on $z>0$ with $Y(0)=0$. Solving it by shooting (with the
Frobenius data~\eqref{eq:frob} for the start) we find a growing mode
$e^{\sigma t}$ ($\Omega^2=-\sigma^2$) for $1<a/2M\lesssim1.28$ and none for larger $a$
(Table~\ref{tab:main}, Fig.~\ref{fig:rates}).

% ---------------------------------------
\subsection{The black-hole branch}
% ---------------------------------------

For $a<2M$ the region $x>x_h$ is static, $\beta'=x/r^2>0$ there, and there is no wall:
both gauges are applicable and equivalent, Eq.~\eqref{eq:Y-B} being a regular change of
variables. Near the horizon $A\simeq A'_h(x-x_h)$, $A'_h=x_h/(4M^2)$, $z\to-\infty$,
and a growing mode must behave as
$Y\sim e^{\sigma z}\sim(x-x_h)^{\sigma/A'_h}$. Since $\Veff<0$ outside a small vicinity
of the horizon (Fig.~\ref{fig:pot}b), bound states with $\Omega^2<0$ are expected. We
find a leading growing mode for all $0<a<2M$, with $\Omega_0^2\to-0.03652\,M^{-2}$
($\sigma_0=0.1911/M$) as $a\to0$, decreasing in magnitude towards the extremal limit 
with $M\sigma_0\to 0.10121$ as $a/(2M)\to 1^-$
(Table~\ref{tab:main}). For $a/2M\gtrsim0.3$ we also find a second, extremely shallow mode
($\sigma_1\approx1\times10^{-5}/M$ at $a/2M=0.3$, $1\times10^{-4}/M$ at $a/2M=0.5$, 
$2.3\times10^{-3}/M$ at $a/2M=0.9$); for smaller $a$ it is too shallow to be resolved, if it exists.
The potential~\eqref{eq:Veff-SV} has a finite limit $\Veff=-6MA/x^3$ as $a\to 0$, which is
why $\sigma_0$ becomes independent of $a$ for small $a$ even though the scalar field
itself degenerates. All black holes of the family are therefore linearly unstable; 
the would-be throat is now hidden beyond the horizon in the form of a black bounce and 
plays no role.

% ===============================================================
\section{Darboux transformations and isospectrality}
\label{sec:darboux}
% ===============================================================

\subsection{Isospectrality theorem}

Let $H=-\dd^2/\dd z^2+V(z)$ and let $Y_s$ be a real solution of $HY_s=E_sY_s$, the
\emph{seed}, with no zeros on the interval considered (except where it is singular).
With $W\equiv Y_s'/Y_s$ and $\mathcal A=\dd/\dd z-W$, $\mathcal A^\dagger=-\dd/\dd z-W$,
\begin{align}
  &H=\mathcal A^\dagger\mathcal A+E_s,\qquad
  \widetilde H\equiv\mathcal A\mathcal A^\dagger+E_s=-\frac{\dd^2}{\dd z^2}+\Vt,\nonumber\\
  &\Vt=V-2W'=2W^2-V+2E_s .
  \label{eq:darboux}
\end{align}
The intertwining relations $\widetilde H\mathcal A=\mathcal AH$ and
$H\mathcal A^\dagger=\mathcal A^\dagger\widetilde H$ imply:

\emph{Theorem.} If $W$ is smooth on the whole axis (i.e., $Y_s$ is nodeless) and
$W\to\pm\kappa_s\ne0$ at $z\to\pm\infty$ such that $\mathcal A$ maps decaying solutions
into decaying ones, then for every $E\ne E_s$ the map $Y\mapsto\widetilde Y=\mathcal AY$ is
a bijection between the solutions of $HY=EY$ satisfying the physical boundary conditions
and the normalizable solutions of $\widetilde H\widetilde Y=E\widetilde Y$, with the inverse
$Y=\mathcal A^\dagger\widetilde Y/(E-E_s)$. Hence the spectra of the two problems
coincide, except possibly for $E_s$, which is an eigenvalue of $\widetilde H$ if and only if
$1/Y_s$ is normalizable.

The theorem \cite{Darboux1882} (see also, e.g., 
\cite{CooperKhareSukhatme1995,Glampedakis2017}) is the basis of the S-deformation
method~\cite{BronnikovFabrisZhidenko2011,BronnikovKonoplyaZhidenko2012,Bronnikov2018Particles}
and of the intertwining operator of~\cite{GGSLinear2009}. Its essential condition is the
absence of zeros of $Y_s$: at a zero of $Y_s$, $W$ has a pole and $\Vt$ acquires a new wall $2/(z-z_n)^2$.

% --------------------------------------
\subsection{Removing the wall}
% --------------------------------------

Take $Y_s$ from the singular branch~\eqref{eq:frob} (with $c_2=0$, so that $Y_s$ is odd
in $z$). Then $W=-1/z+(E_s-w_0)z+O(z^3)$, the $1/z^2$ terms in $\Vt$ cancel, and
\begin{equation}
  \Vt(0)=3w_0-2E_s .
  \label{eq:Vt0}
\end{equation}
The partner potential is regular and even, and the problem for $\widetilde H$ is a standard
one on the whole axis $z\in\mathbb R$. Near the throat,
$\widetilde Y=\mathcal AY$ is even for $Y\sim Y_{\rm sing}$ and odd for $Y\sim Y_{\rm reg}$:
\emph{the even sector of $\widetilde H$ describes the throat-radius modes, and its odd sector
reproduces the naive spectrum} (plus the level $E_s$, whose eigenfunction $1/Y_s\sim z$ is
odd and normalizable when $Y_s$ grows at infinity). In terms of the gauge-invariant
$\mathcal B$, $\widetilde Y=Y_s\,(\mathcal B/\mathcal B_s)_z$, a form that we will meet again in
Sec.~\ref{sec:cps}.

% -------------------------------------------
\subsection{The Ellis limit}
% -------------------------------------------

\begin{figure*}[t]
\centering
\includegraphics[width=0.96\textwidth]{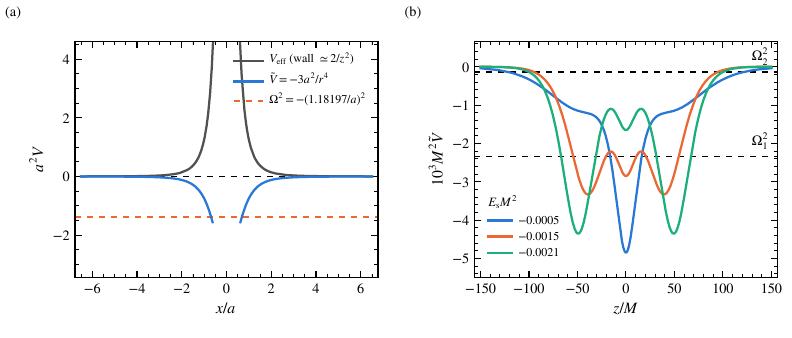}
\caption{Darboux partner potentials. (a) Ellis limit: $\Veff$ with the wall (gray) and its
partner~\eqref{eq:Vt-Ellis} (blue) for the static seed $Y_s=r/x$; the dashed line is the
throat-mode eigenvalue. (b) $a/2M=10$: partners built from three nodeless seeds with
different $E_s$ in the admissible window. The potentials differ visibly, yet their
even-sector eigenvalues (dashed) coincide to $10^{-7}$ (Table~\ref{tab:iso}).}
\label{fig:darboux}
\end{figure*}

For $M=0$ the static seed is known in closed form: $\mathcal B_s=$ const, i.e.,
$Y_s=r/x$, which corresponds to the variation of the throat radius $a$ within the
family of Ellis solutions. It is nodeless, $W=-a^2/(xr^2)$, and
\begin{equation}
  \Vt=-\frac{3a^2}{(x^2+a^2)^2}\qquad (M=0,\ E_s=0),
  \label{eq:Vt-Ellis}
\end{equation}
a regular potential well, see Fig.~\ref{fig:darboux}(a), which reproduces the results
of~\cite{GGSLinear2009,CPS2020}. It has a single bound state, which we find at
$\Omega^2=-(1.18197/a)^2$, an even (throat) mode; the odd sector is empty, in agreement
with the absence of naive instabilities for the Ellis wormhole.

% --------------------------------------------------------- fig 2

% -----------------------------------------------------------
\subsection{SV wormholes: an obstruction for the static seed}
% -----------------------------------------------------------

For $M>0$ the static seed ($E_s=0$, $c_2=0$) always has a zero. The reason is the
attractive tail $\Veff\simeq-6M/x^3$: at large $z$ the static solution behaves as
$k_1+k_2 z$, and the tail bends it towards zero. The location $z_n$ of the zero is
\begin{center}\small
\begin{tabular}{lccccccc}
\toprule
$a/2M$ & 1.05 & 1.2 & 1.5 & 2 & 3 & 5 & 15 \\
$z_n/M$\ &\ 17.9\ \ &\ 10.3\ \ &\ 9.39\ \ &\ 11.3\ \ &\ 17.7\ \ &\ 35.7\ \ &\ 211 \\
\bottomrule
\end{tabular}
\end{center}
(for $a/2M=1.05$ there is a second zero at $z\approx333M$). Hence the transformation
with the static seed removes the wall at $z=0$ but instead creates a new one at some $z = z_n$.
Note that an ad hoc construction, such as using a wall-free combination of the partner and original 
potentials inside a window around the throat, is not an intertwining transformation and is not 
isospectral; results obtained in this way cannot be trusted.

The proper way is to use the remaining freedom: the seed energy $E_s$ and the constant
$c_2$. For $E_s<w_0$, a zero appears near the throat (by~\eqref{eq:frob}, at
$z^2\approx2/(w_0-E_s)$), while for $E_s$ close to zero the far zero persists; a nodeless
seed must lie in between. A systematic scan of $E_s\in(1.3w_0,0)$ and of $c_2$ shows:
\begin{itemize}
\item for $a/2M\gtrsim4.98$, nodeless seeds exist in a window inside $(w_0,0)$, which
numerically coincides with the interval $-\sigma_1^2<E_s<-\sigma_2^2$ between the two throat
eigenvalues found below; e.g., $-0.00370<E_s M^2<-0.00310$ at $a/2M=5$ and
$-0.00235<E_sM^2<-0.00014$ at $a/2M=10$;
\item for $a/2M\lesssim4.98$, we found no nodeless singular-branch seed for any
$E_s\le0$ and any $c_2$ (an admixture $c_2\neq0$ removes the zero on one side of the
throat only at the price of creating one on the other side).
\end{itemize}

% ----------------------------------------------
\subsection{Results of the Darboux method}
% ----------------------------------------------

Where nodeless seeds exist, $\Vt$ is regular and $\widetilde H$ is self-adjoint. We computed
its spectrum by diagonalizing a second-order finite-difference approximation on
$z\in[0,Z]$ (Neumann/Dirichlet conditions at $z=0$ for the even/odd sector, $Z$ up to
$6000M$). The even sector has exactly two negative eigenvalues; the odd sector contains
only the auxiliary level $E_s$, as predicted by the theorem (there are no naive modes
for $a/2M\gtrsim1.28$). Table~\ref{tab:iso} shows that partner potentials built from
different seeds give the same eigenvalues to $10^{-7}$, a direct check of
isospectrality, and that they coincide with the CPS results of the next section
to within a few units of $10^{-7}$.

\begin{table}[t]
\caption{Even-sector (throat) eigenvalues $\sigma=\sqrt{-\Omega^2}$ (units $M=1$) from
Darboux partners with different seeds, compared with the CPS equation~\eqref{eq:CPS}.}
\label{tab:iso}
\small
\begin{tabular}{@{}lccc@{}}
\toprule
$a/2M$ & \ seed $E_s$ & $\sigma_1$ & $\sigma_2$\\
\midrule\
6  & $-0.003$  &\ 0.0657430 &\ 0.0326763\\
   & CPS       &\ 0.0657430 &\ 0.0326759\\
10 & $-0.0005$ &\ 0.0484547 &\ 0.0116910\\
   & $-0.0015$ &\ 0.0484546 &\ 0.0116911\\
   & $-0.0021$ &\ 0.0484546 &\ 0.0116911\\
   & CPS       &\ 0.0484546 &\ 0.0116910\\
15 & $-0.0005$ &\ 0.0348008 &\ 0.0054070\\
   & CPS       &\ 0.0348008 &\ 0.0054070\\
\bottomrule
\end{tabular}
\end{table}

The disappearance of nodeless seeds at $a/2M\approx4.98$ (EP1) is not a technical accident:
the window $(-\sigma_1^2,-\sigma_2^2)$ shrinks to zero there.
At exactly this value the two real throat eigenvalues merge and, for smaller $a$, turn
into a complex-conjugate pair of $\Omega^2$ (Sec.~\ref{sec:cps}). A real regular $\Vt$
defines a self-adjoint operator, which cannot have complex eigenvalues; therefore no
isospectral regular partner of this type can exist for $a/2M<4.98$. The Darboux method
thus correctly signals its own limits, but to cover the whole wormhole branch we need a
method that does not rely on a seed.

%======================================================================
\section{The CPS method: a single regular equation}
\label{sec:cps}
%======================================================================

\subsection{Gauge-invariant variables and the master equation}

Cremona, Pizzocchero and Sarbach~\cite{CPS2020} formulated spherical perturbations of
wormholes supported by a self-interacting phantom field in terms of gauge-invariant
combinations of the metric and scalar perturbations, constructed so that no division by
$\beta'$ occurs, and hence nothing singular appears at the throat. In the gauge
$\delta\psi=0$, which is always attainable since $\psi'\neq0$, their invariants reduce
to the lapse, radial-metric and areal-radius perturbations, $\delta\gamma$,
$\delta\alpha$ and $\delta r=r\delta\beta$; in particular, their variable
$G=\delta r/r^2$ equals $\mathcal B/r$. Their momentum constraint is our
Eq.~\eqref{eq:C01} with $\delta\psi=0$, and their Hamiltonian constraint is
Eq.~\eqref{eq:C00}.

We now extend the construction to the NED-augmented source. As noted in
Sec.~\ref{sec:pert}, NED enters only through $\frac12 L_\beta\delta\beta$ in the
diagonal components; it gives no momentum flux and does not couple to the scalar
equation, so that the CPS constraints keep their pure-scalar form, while the dynamical
equation acquires the $L_\beta$ term. Eliminating $\delta\alpha$ and $\delta\gamma$ as in
Sec.~\ref{sec:pert}, we obtain a \emph{single equation for the gauge-invariant
$\mathcal B$} (equivalently, for the CPS variable $G$). Writing
$Y=g\mathcal B$ with $g\equiv-r\psi'/\beta'$ in Eq.~\eqref{eq:master} and keeping
$\mathcal B$ as the unknown, we get, for $\mathcal B\propto e^{i\Omega t}$,
\begin{align}
  &\frac{\dd^2\mathcal B}{\dd z^2}-2S(z)\frac{\dd\mathcal B}{\dd z}
  +\big[\Omega^2-\VC(z)\big]\mathcal B=0,\nonumber\\
  &S=-\frac{g_z}{g},\qquad \VC=\Veff-\frac{g_{zz}}{g},
  \label{eq:CPS}
\end{align}
or, in self-adjoint form, $(g^2\mathcal B_z)_z+g^2(\Omega^2-\VC)\mathcal B=0$.
For the SV background, $g=-ar/x$ and the algebra collapses: the whole numerator $N(x)$
of~\eqref{eq:Veff-SV} cancels except for one term, and
\begin{equation}
  \boxed{\ \VC=-\frac{6MA(x)\,x^2}{r^5},\qquad S=\frac{a^2A(x)}{x\,r^2}\ }
  \label{eq:VC}
\end{equation}

Equation~\eqref{eq:CPS} has several useful properties.
(i) The potential $\VC$ is regular, even in $x$, nonpositive for wormholes, vanishes at
the throat and at infinity, and is identically zero in the Ellis limit $M=0$
(Fig.~\ref{fig:pot}a). It coincides with $\Veff$ in the limit $a\to0$.
(ii) The wall has been converted into the first-order term $-2S\mathcal B_z$ with
$S\simeq1/z$; the leading part $\mathcal B_{zz}-(2/z)\mathcal B_z$ is the radial
Laplacian in ``$d=-1$ dimensions''. Its indicial exponents are $0$ and $3$, both
solutions are smooth [Eq.~\eqref{eq:B-loc}], and no boundary condition is needed at the
throat beyond smoothness. Both sectors, even and odd, are thus treated on an equal
footing.
(iii) The equation is independent of $q$ and contains the scalar and NED sources only
through the background.
(iv) In the original CPS construction one next uses a static solution $(F_0,G_0)$ with
$G_0\neq0$ to pass to a decoupled variable
$\tilde\chi=F-(F_0/G_0)G$~\cite{CPS2020}. In our language this is exactly the Darboux
variable $\widetilde Y=Y_s(\mathcal B/\mathcal B_s)_z$ of Sec.~\ref{sec:darboux} with a static
seed. In the Ellis limit $\mathcal B_s={\rm const}$ and $\widetilde Y\propto(r/x)\mathcal B_x$,
which leads to Eq.~\eqref{eq:Vt-Ellis}. For $M>0$ the static solution has a zero (see the
table in Sec.~\ref{sec:darboux}), so this last step fails. We therefore do not take it,
and solve Eq.~\eqref{eq:CPS} directly.

\subsection{Numerical methods}

\emph{Shooting.} For each sector we construct the power series of $\mathcal B$ at
$x=0$ up to $x^{15}$ (even: $\mathcal B=1+\ldots$, with the free coefficient of $x^3$ set
to zero; odd: $\mathcal B=x^3+\ldots$), start the integration at
$x_0=\min\{0.05a,\,0.05A(0)/|\kappa|\}$, and integrate Eq.~\eqref{eq:CPS} outward with
an eighth-order Runge--Kutta method (relative tolerance $10^{-12}$) to
$z_{\max}\geq\max\{30/{\rm Re}\,\kappa,\,100M+20a\}$. With $\Omega^2=-\kappa^2$,
${\rm Re}\,\kappa>0$, the solution behaves as $c_+e^{\kappa z}+c_-e^{-\kappa z}$ at large
$z$, and eigenvalues are the zeros of $c_+(\kappa)$. Since the even sector is
\emph{not} self-adjoint in the variable $Y$ (the norm $\int Y^2\dd z$ diverges), we do not
restrict $\kappa$ to real values but search for zeros of $c_+$ in the complex half-plane
${\rm Re}\,\kappa>0$ by the secant method, tracking each root as $a$ varies. A complex
$\kappa$ describes a growing oscillation $\propto e^{{\rm Re}\kappa\,t}\cos({\rm
Im}\kappa\,t)$, accompanied by its complex conjugate.

\emph{Time domain.} As a completely independent check, we evolve smooth even initial
data in time. A direct discretization of the term $-2S\mathcal B_z$ with $S\simeq 1/z$ is
inaccurate, so we introduce, by analogy with the dimension-raising trick for radial wave
equations, the auxiliary even variable $w=\mathcal B_x/x$. Equation~\eqref{eq:CPS} then
becomes the regular system
\begin{align}
  \ddot{\mathcal B}&=A^2x\,w'+c\,w-\VC\mathcal B, \nonumber\\
  \ddot w&=A^2w''+D\,w'+\Big(\frac{c'}{x}-\VC\Big)w-\frac{\VC'}{x}\,\mathcal B,
  \label{eq:td}
\end{align}
with $c=A^2(x^2-a^2)/r^2+AA'x$ and $D=3AA'+2A^2x/r^2$, all coefficients being smooth
at $x=0$. We solve~\eqref{eq:td} with second-order finite differences on a cell-centered
grid and leapfrog time stepping, with Gaussian initial data at rest and an outer boundary
far enough to avoid reflections, and extract the dominant $\kappa$ from the late-time
signal $\mathcal B(t,0)$ by linear prediction (Prony's method). In the Ellis case this gives
$\kappa a=1.18203$ at grid step $h=0.02a$ and $1.18220$ at $h=0.04a$, converging to
the eigenvalue $1.18197$ of~\eqref{eq:Vt-Ellis}.

% ------------------------------
\subsection{Results}
% ------------------------------

\begin{figure*}[t]
\centering
\includegraphics[width=0.96\textwidth]{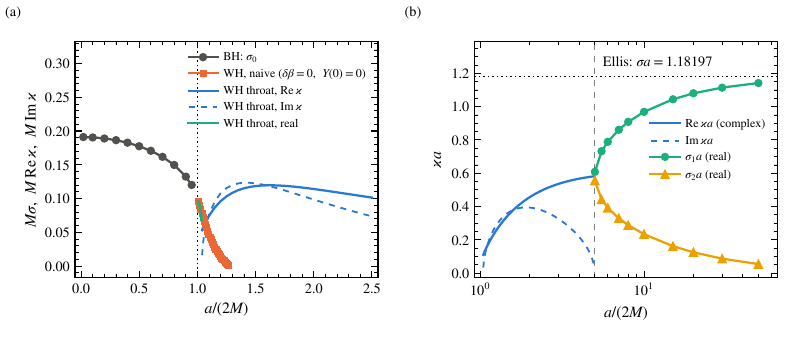}
\caption{(a) Growth rates for all branches ($M=1$): black holes (gray), naive odd-sector
wormhole modes (orange), throat modes (blue: ${\rm Re}\,\kappa$ solid, ${\rm Im}\,\kappa$
dashed; green: real near extremality). (b) Throat modes on a logarithmic scale in units
of $1/a$: the complex pair (blue) splits at $a/2M\approx4.98$ (vertical line) into two
real modes; $\sigma_1a$ approaches the Ellis value.}
\label{fig:rates}
\end{figure*}

\begin{figure*}[t]
\centering
\includegraphics[width=0.96\textwidth]{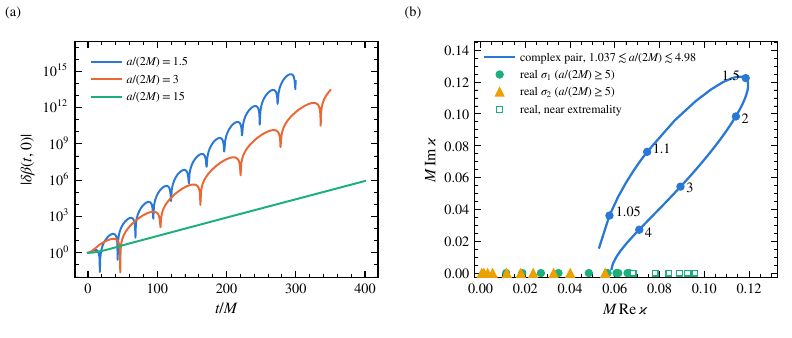}
\caption{(a) Time-domain evolution of~\eqref{eq:td}: $|\delta\beta|$ at the throat for
$a/2M=1.5$, $3$ (oscillatory growth) and $15$ (monotonic growth). (b) Trajectory of
the throat-mode $\kappa$ in the complex plane as $a/2M$ varies (labels); real modes are
shown on the horizontal axis (circles and triangles: $a/2M\geq5$; squares: near
extremality).}
\label{fig:td}
\end{figure*}

The results of all methods are collected in Table~\ref{tab:main} and
Figs.~\ref{fig:rates} and~\ref{fig:td}.

\emph{Odd sector.} The shooting results for $\mathcal B_{\rm odd}$ coincide with those of the
naive $\delta\beta=0$ problem to all digits shown, as they must.

\emph{Even (throat) sector.} A throat mode exists for \emph{every} $a>2M$. Its behavior
has three regimes:
\begin{itemize}
\item $a/2M\gtrsim4.98$: two real growing modes, $\sigma_1>\sigma_2$. As
$a/2M\to\infty$, $\sigma_1a\to1.18197$, the Ellis value, and $\sigma_2a\to0$; the second
mode is created by the attractive tail $\VC\propto-M$ and is absent for $M=0$.
\item $1.037\lesssim a/2M\lesssim4.98$: the two modes merge at $a/2M\approx4.98$ and form
a complex-conjugate pair $\kappa={\rm Re}\,\kappa\pm i\,{\rm Im}\,\kappa$ with
${\rm Re}\,\kappa$ between $0.05$ and $0.12\,M^{-1}$ and a maximum of
${\rm Re}\,\kappa\approx0.1196/M$ near $a/2M\approx1.6$
(Fig.~\ref{fig:td}b).
\item 
$1<a/2M\lesssim1.037$: 
the complex pair has returned to the real axis at EP2,
while two weaker real modes merge at EP3 and form another
complex-conjugate pair for smaller values of $a/(2M)$;
the leading real mode grows to $\sigma\approx0.096/M$
at $a/2M=1.01$, close to the naive odd mode.
An additional weak real mode also persists beyond
EP2. The detailed mode structure is discussed below.
\end{itemize}

\emph{Cross-checks.} Where nodeless seeds exist ($a/2M\gtrsim4.98$), the Darboux
spectra agree with the CPS results to a few units of $10^{-7}$ (Table~\ref{tab:iso}). The time-domain
evolutions reproduce the leading real and complex modes to five digits, and the weak
second real mode at $a/2M=15$ to about $1\%$
(Table~\ref{tab:td}); in Fig.~\ref{fig:td}(a), the throat value of $\delta\beta$ grows
exponentially with superimposed oscillations for $a/2M=1.5$ and $3$ and grows
monotonically for $a/2M=15$. Close to extremality ($a/2M\lesssim1.05$) the time-domain
scheme converges slowly because $A(0)\to0$, and we rely on shooting there.

\emph{Black holes.} For $a<2M$ the transformation $Y=g\mathcal B$ is regular outside the
horizon, so Eq.~\eqref{eq:CPS} and the master equation~\eqref{eq:master} give identical
spectra; the black-hole results of Sec.~\ref{sec:pert} stand.

\begin{table}[t]
\caption{Growth rates (units $M=1$). BH: $\sigma_0$ for $a<2M$. WH: naive (odd-sector)
$\sigma$ and the throat (even-sector) modes $\kappa$ of Eq.~\eqref{eq:CPS}; a complex
entry stands for a pair $\kappa,\bar\kappa$, two real entries for two modes. ``---'':
no growing mode.}
\label{tab:main}
\small
\begin{tabular}{@{}lccc@{}}
\toprule
 & $a/2M$ & naive $\sigma$ (odd) & throat $\kappa$ (even)\\
\midrule
BH & 0.02 & \multicolumn{2}{c}{$\sigma_0=0.19109$, $\Omega_0^2=-0.03652$}\\
BH & 0.3  & \multicolumn{2}{c}{$\sigma_0=0.18643$, $\Omega_0^2=-0.03476$}\\
BH & 0.5  & \multicolumn{2}{c}{$\sigma_0=0.17745$, $\Omega_0^2=-0.03149$}\\
BH & 0.7  & \multicolumn{2}{c}{$\sigma_0=0.16176$, $\Omega_0^2=-0.02617$}\\
BH & 0.9  & \multicolumn{2}{c}{$\sigma_0=0.13257$, $\Omega_0^2=-0.01758$}\\
BH & 0.95 & \multicolumn{2}{c}{$\sigma_0=0.12004$, $\Omega_0^2=-0.01441$}\\
\midrule
WH & 1.01 & 0.09588 & 0.09582\\
WH & 1.02 & 0.08999 & 0.08885\\
WH & 1.05 & 0.07225 & $0.05763\pm0.03619i$;\ 0.00350\\
WH & 1.1  & 0.04715 & $0.07445\pm0.07616i$\\
WH & 1.2  & 0.01419 & $0.09617\pm0.10803i$\\
WH & 1.5  & ---     & $0.11847\pm0.12242i$\\
WH & 2    & ---     & $0.11412\pm0.09847i$\\
WH & 3    & ---     & $0.08938\pm0.05428i$\\
WH & 4    & ---     & $0.07090\pm0.02730i$\\
WH & 5    & ---     & 0.06081;\ 0.05564\\
WH & 6    & ---     & 0.06574;\ 0.03268\\
WH & 10   & ---     & 0.04845;\ 0.01169\\
WH & 15   & ---     & 0.03480;\ 0.00541\\
WH & 30   & ---     & 0.01857;\ 0.00145\\
WH & 50   & ---     & 0.01142;\ 0.00054\\
\midrule
Ellis & $\infty$ & --- & $\kappa a=1.18197$\\
\bottomrule
\end{tabular}
\end{table}

\begin{table}[t]
\caption{Throat modes: shooting for Eq.~\eqref{eq:CPS} vs.\ time-domain evolution
of~\eqref{eq:td} (Prony fit of $\mathcal B(t,0)$), units $M=1$.}
\label{tab:td}
\small
\begin{tabular}{@{}lcc@{}}
\toprule
$a/2M$ & shooting & time domain\\
\midrule
1.5 & $0.118473\pm0.122425i$ & $0.11847\pm0.12242i$\\
2   & $0.114117\pm0.098472i$ & $0.11412\pm0.09847i$\\
3   & $0.089377\pm0.054283i$ & $0.08938\pm0.05428i$\\
15  & $0.034801$;\ $0.005407$ & $0.03480$;\ $0.00534$\\
Ellis ($\kappa a$) & 1.18197 & 1.18203\\
\bottomrule
\end{tabular}
\end{table}
% -------------------------------------------------------- fig 3
% --------------------------------------------------------- fig 4
% -------------------------------------------------------- fig 5
\begin{figure*}[t]
\centering
\includegraphics[width=0.96\textwidth]{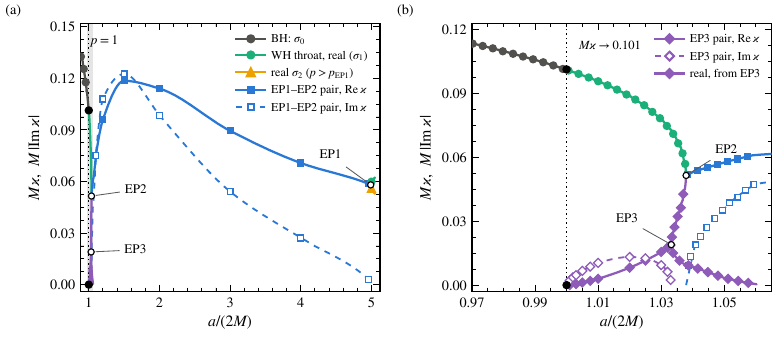}
\caption{Unstable eigenvalues of the Simpson--Visser
black holes and wormholes as functions of $p=a/(2M)$.
(a) Overall spectrum, showing the real and
complex unstable modes and the three exceptional points EP1, EP2, and EP3.
(b) Enlargement of the near-extremal region,
showing the three real branches between EP2 and EP3,
the continuation of the weak real mode beyond EP2,
and the complex-conjugate pair below EP3. Solid curves represent real growth rates,
while dashed curves indicate the absolute values of the imaginary parts.
The leading black-hole and wormhole branches
are joined at $p=1$, where their limiting growth
rates coincide at $M\kappa\simeq0.10121$.
The regularity of the perturbations at the extremal horizon has not been established.
}
\label{fig:EPs}
\end{figure*}
% ------------------------------------------------------------------

% ---------------------------------------------------------------------------------
\subsection{Exceptional points and the near-extremal limit}\label{sec:sv_eps}
% ---------------------------------------------------------------------------------

A more detailed analysis of the unstable spectrum reveals three exceptional points,
denoted by EP1, EP2, and EP3, at $p=a/(2M)\simeq4.98477$, $1.03773$, and
$1.03318$, respectively. At these points, two eigenvalues and their corresponding
eigenfunctions coalesce. The numerical results also exhibit the characteristic square-root
behavior of the eigenvalues in their vicinity \cite{HeissEP2012}.
The resulting mode structure is shown in Fig.~\ref{fig:EPs}.

As $p$ decreases, the two real unstable
modes merge at EP1 and form a complex-conjugate
pair. At EP2, this pair returns to the real
axis. Between EP2 and EP3, three distinct
real unstable modes are present. The upper
and middle branches merge at EP2, whereas
the middle and lower branches merge at EP3.
Thus, the middle branch connects EP2 and EP3,
while the lower real branch continues beyond
EP2. Below EP3, the two weaker modes form
another complex-conjugate pair, whose real
and imaginary parts both approach zero as $p\to 1^+$.

The leading real wormhole mode behaves differently. Its growth rate approaches
$M\kappa\simeq0.10121$ as $p\to1^+$. The leading black-hole instability approaches
the same value as $p\to1^-$. We have therefore joined the corresponding curves at $p=1$
in Fig.~\ref{fig:EPs}, illustrating the continuity of the limiting eigenvalues.
This does not establish the regularity of the perturbations at the extremal horizon,
which requires a separate analysis.

% -------------------------------------------
\subsection{Comparison and inferences}
% -------------------------------------------

Comparing the methods, we can draw the following conclusions.

1. The naive $\delta\beta=0$ analysis is correct for what it describes, the odd sector,
but misses the even sector entirely. It suggests a stable window $a/2M\gtrsim1.28$, which does
not exist.

2. The Darboux transformation is exact and isospectral whenever a nodeless seed exists;
for SV wormholes this happens only for $a/2M\gtrsim4.98$, where the throat spectrum is
real. Its failure for smaller $a$ reflects the complex spectrum, not a technical
limitation that could be removed by a cleverer seed of the same type.

3. The CPS-type equation~\eqref{eq:CPS} works for all $a \ne 2M$ and contains both sectors.
Its potential $\VC=-6MAx^2/r^5$ is regular and remarkably simple, and the time-domain
evolution confirms its spectrum.

4. All three methods agree wherever they overlap: Ellis limit, odd sector, and
$a/2M\gtrsim4.98$. The SV black holes are unstable, with a nearly $a$-independent rate
$\sigma_0\approx0.19/M$ for $a\ll2M$, and the SV wormholes are unstable for all $a>2M$.
The dominant instability is the throat mode, except very close to extremality, where the
naive odd mode is comparable to it, and its $e$-folding time $1/{\rm Re}\,\kappa$ ranges
from about $8M$ to $20M$ for $a/2M\lesssim 5$ and grows as $a/1.18$ for very thick wormholes.

%======================================================================
\section{Discussion and outlook}
\label{sec:disc}
%======================================================================

We have analyzed radial perturbations of the SV black-bounce space-times supported by a
phantom scalar with a potential and magnetic NED~\cite{BronnikovWalia2022}. The NED
field, being a fixed function of $r$, does not affect the perturbation equations beyond
the background, so the results are independent of the magnetic charge. The main result
is that \emph{the nonextremal family is linearly unstable}: the black holes ($a<2M$) because of a
bound state of the master potential outside the horizon, and the wormholes ($a>2M$)
because of the throat-radius mode, throughout the corresponding parameter ranges.

The throat mode is invisible in the gauge $\delta\beta=0$ under the naive boundary
condition. We have shown that this is a matter of admissible solutions rather than of the
potential: in the gauge $\delta\psi=0$, the normal form of the equations is the same,
wall included, but the throat-radius perturbations are perfectly regular there. We have
then treated the throat in two ways. The Darboux transformation, with its isospectrality
theorem, gives an exact regular partner when a nodeless seed exists; the static seed
used for the Ellis wormhole always acquires a zero for $M>0$, and nodeless seeds exist
only for $a/2M\gtrsim4.98$. The CPS-type gauge-invariant equation with the regular
potential $\VC=-6MAx^2/r^5$ requires no seed and works for all $a \ne 2M$. The two methods agree where both apply, and the time-domain evolution confirms the results.

A notable feature is that for $1.037\lesssim a/2M\lesssim4.98$ the throat modes come as
complex-conjugate pairs of $\Omega^2$, i.e., the instability is oscillatory. This is
possible because the throat sector is not self-adjoint with respect to the natural norm
of the master equation, and the energy of phantom perturbations is not positive
definite. Oscillatory growth was found for electrically charged ghost wormholes
in~\cite{GGSCharged2009}; here it occurs without electric charge and is caused by the
mass term, i.e., by the potential $V(\psi)$ and the NED stress in the background. In
the Ellis limit the spectrum becomes real, and we recover the known growth rate
$\kappa a=1.18197$.

A more detailed spectral analysis reveals three exceptional points, extending the 
exceptional-point structure recently found for rotating five-dimensional
Ellis--Bronnikov wormholes~\cite{EB5DEP2026}.
In particular, two near-extremal exceptional points
are connected by a common real eigenvalue branch.
As $a/(2M)\to 1^+$, the weak complex-conjugate pair
approaches zero, whereas the leading real wormhole
mode approaches a finite growth rate.
Remarkably, this limiting growth rate,
$M\kappa\simeq 0.10121$, coincides with that of
the leading black-hole instability as $a/(2M)\to 1^-$.

Our conclusions confirm, for this family, the general expectation that throats supported
by phantom scalars are unstable~\cite{BronnikovKonoplyaZhidenko2012,Bronnikov2018Particles}.
They also show that a ``thick'' throat ($a\gg 2M$) does not help: the growth rate decreases
only as $1/a$. Since a stability analysis restricts the parameter values that can be
realized as stationary objects, our results, together with shadow and light-ring
studies~\cite{AlencarEtAl2026Shadows}, indicate that the SV geometry with this field source cannot describe long-lived astrophysical objects.

Several extensions are natural. (i) A nonzero electric charge couples NED to the
perturbations through $\delta\mathcal F$ and may affect the instability, as
in~\cite{GGSCharged2009}. 
(ii) Although the leading black-hole and wormhole
growth rates approach the same value as
$a/(2M)\to 1$ from either side, the regularity
of perturbations at the degenerate horizon
of the extremal configuration $a=2M$
remains to be established.
(iii) Nonradial perturbations ($\ell\ge1$) and
(iv) the nonlinear fate of the instability (collapse to a black hole or
expansion~\cite{GGSNonlinear2009}) remain open. (v) The single-equation approach of
Sec.~\ref{sec:cps}, with the decomposition $\Veff=\VC+g_{zz}/g$, applies to any
phantom-scalar wormhole with a monotonic scalar field, and may be useful for long-throat geometries \cite{BronnikovKorolyov2017}, 
for which potential walls can only be cured by multiple Darboux transformations.

%import math
%print(math.sqrt(16))
%\end{python}

% ============================
\begin{acknowledgments}
The research of K.~B.\ was supported in part by the Ministry of Science and Higher
Education of the Russian Federation, Project FSWU-2026-0010. R.~I.\ gratefully
acknowledges support from the Ministry of Innovative Development of the Republic of
Uzbekistan, Project No.\ FZ-20200929385.
\end{acknowledgments}

% ============================

\end{document}